\documentclass[12pt,final,1p,times]{elsarticle}

\usepackage{pifont}
\usepackage{natbib}
\usepackage{geometry}
\usepackage{graphicx}
\usepackage{txfonts}
\usepackage{subfigure}
\usepackage{wrapfig}
\usepackage{color}
\usepackage{booktabs}
\usepackage{caption}
\usepackage{amsmath}
\usepackage{amssymb}
\usepackage{url}
\usepackage{colortbl}
\usepackage{multirow}
\usepackage{fnbreak}
\usepackage{helvet}
\usepackage[font=bf]{caption}

\graphicspath{{figs/}}

\definecolor{Gray}{gray}{0.7}

\begin{document}

\begin{frontmatter}
\begin{center}
\textbf{3\textsuperscript{rd} IAA Conference on Space Situational Awareness (ICSSA)} \\ \vspace{0.1in} \textbf{GMV, Madrid, Spain} \\ \vspace{0.4in} \textbf{IAA-ICSSA: 4-6/04/2022} \\
\textbf{Machine learning to predict the solar flux and geomagnetic indices to model density and Drag in Satellites}
\end{center}


\author{\textbf{S. Aljbaae}$^{(1,2)}$}
\ead{safwan.aljbaae@gmail.com}
\author[fml3]{\textbf{J. Murcia-Piñeros}}
\author[fml1]{\textbf{A. F. B. A. Prado}}
\author[fml3]{\textbf{R. V. Moraes}} 
\author[fml4]{\textbf{V. Carruba}}
\author[fml1]{\textbf{G. A. Caritá}}


\address[fml1]{\normalsize Division of Post-Graduate Studies, INPE, C.P. 515, 12227-310 S\~ao Jos\'e dos Campos, SP, Brazil.}
\address[fml2]{\normalsize 2RP Net - Data-driven Company, Av. Paulista, 1159, sala 1511 - Bela Vista, SP, Brazil.}
\address[fml3]{\normalsize São Paulo Federal University (UNEFESP), Institute of Science and Technology (ICT), São José dos Campos, SP, 12247-014, Brazil.}
\address[fml4]{\normalsize São Paulo State University (UNESP), School of Natural Sciences and Engineering, Guaratinguetá, SP, 12516-410, Brazil.}

\end{frontmatter}

\begin{flushleft}
\end{flushleft}

\section*{}
   In recent years (2000-2021), human-space activities have been increasing faster than ever. More than 36000 Earth' orbiting objects, all larger than 10 cm, in orbit around the Earth, are currently tracked by the European Space Agency (ESA)\footnote{\href{https://www.esa.int/Safety_Security/Space_Debris/Space_debris_by_the_numbers}{https://www.esa.int/Safety\_Security/Space\_Debris/Space\_debris\_by\_the\_numbers}}. Around 70\% of all cataloged objects are in Low-Earth Orbit (LEO). Aerodynamic drag provides one of the main sources of perturbations in this population, gradually decreasing the semi-major axis and period of the LEO satellites. Usually, an empirical atmosphere model as a function of solar radio flux and geomagnetic data is used to calculate the orbital decay and lifetimes of LEO satellites. In this respect, a good forecast for the space weather data could be a key tool to improve the model of drag. In this work, we propose using Time Series Forecasting Model to predict the future behavior of the solar flux and to calculate the atmospheric density, to improve the analytical models and reduce the drag uncertainty.

\section{Introduction}\label{sec01_introduction}
   During the last years, an exponential increase in the population of objects in orbit around our planet has been observed, especially in Low Earth Orbits (LEO, \citet{nasa_2021}). That reduces the available operational orbits. It also increases the probability of collisions, which, when happening, results in clouds of debris propagating around the orbits, like for the case of the Iridium 33 and Kosmos 2251 artificial satellites. Moreover, recent activities, like the multiple tests of Anti Satellite Weapons (ASAT), and the new Large Scale Constellations, could increase exponentially the population of objects around Earth in the next few years. If these activities are not controlled and/or regulated, the possible occurrence of a catastrophic scenario is known as the Kessler effect, could limit the space activities and access to orbit for a long period \citep{Kessler_1991}.\\
   
   The environment of the LEO naturally influence the mitigation of artificial objects in orbit, due to the loss of the orbit mechanical energy, which is influenced by the atmospheric-satellite interaction, mathematically modeled as the perturbation caused by drag. Usually, a simplified model for satellites in LEO is used to reduce the computational cost during the propagations, where the main forces that influence the motion are the Keplerian gravity field of the Earth, the perturbation due to the non-sphericity of the central body (J2 and J4 terms of the gravitational perturbation) and the atmospheric drag. Other perturbations, like a third-body (either the Moon or the Sun), Solar Radiation Pressure, tides, and albedo could be negligible at altitudes lower than 400 km \citep{vallado_2007, dellelce_2015}. With the previous considerations, the equation of motion of the satellite moving in LEO in an inertial system, located at the Earth´s center of mass, is written as:
   
   \begin{eqnarray}\label{motion1}
      \ddot{\vec{r}} =-\vec{g_{4 \times 4}} + \vec{a}_{D}
   \end{eqnarray}
   
   \noindent where $g_{4 \times 4}$ represents the Earth´s Gravitational Model (EGM-08) of order $4\times4$, $r$ is the inertial acceleration vector, and $a_{D}$ is the drag acceleration vector, which is acting in the opposite direction of the airflow vector $\vec{V}_{\infty}$. The airflow is the difference between the inertial velocity vectors and the atmospheric velocity due to the Earth´s rotation, including the winds. Several models of drag have been applied to determine the atmosphere-satellite interaction and to reduce the uncertainty \citep{prieto_2014}. The basic drag acceleration model is described as follows
   
   \begin{eqnarray}\label{drag}
      \vec{a}_{D} = -\rho \bigg(\frac{C_{D} A}{2m} \bigg) V_{\infty} \vec{V}_{\infty}
   \end{eqnarray}
   \noindent where, $A$ is the satellite’s mean area normal to its velocity vector, which is a difficult parameter to estimate due to the winds and changes in attitude. $C_{D}$ is the drag coefficient, which is a dimensionless quantity indicating the satellite’s susceptibility to drag forces, and $m$ is the satellite’s mass. The quantity $m/A C_{D}$ is usually called the ballistic coefficient. A satellite with a low ballistic coefficient will be considerably affected by the drag forces. $\rho$ is the atmospheric density, and it is a function of the solar activity, local time, altitude and geographic coordinates. As such, it is a rather difficult parameter to estimate. For more details about modelling the aerodynamic drag, we refer the readers to \citep{vallado_2007, vallado_2014, zhejun_2017}.\\
   
   Due to the satellite geometry, materials, and uncertainly of the attitude, the $C_{D}$ is approximated by a mean value, reported in the scientific literature as 2.2 for satellites in the upper atmosphere in Free Molecular Flow (FMF) \citep{vallado_2007}. With the information of the satellite geometry, attitude and materials, it is possible to implement a high fidelity model of the drag for FMF and/or Rarefied Flow, as presented in \citet{prieto_2014, rafano_2019, tewari_2009}, however, this is out of the scope of this present research. In fact, the main problem for orbital determination and propagation in LEO is the accuracy of the drag perturbation. As shown in Eq. \ref{drag}, the drag model is a function of the atmospheric density and, at the same time, it is a function of the space weather, which is a stochastic effect because of the multiple uncertainties affecting it, like the atmospheric conditions due to the solar and geomagnetic activity or the atmospheric density estimations due to the use of empirical models and the atmospheric dynamics (including winds).\\ 
   
   \citet{vallado_2014} discussed the importance of atmospheric modeling to drag estimation. They presented a detailed description of the uncertainties to model the drag, the differences between the atmospheric models, and the functions used to estimate the solar flux and geomagnetic data. For the solar cycle forecast, \citet{schatten_1987} presents a good analytical approximation, which predicts the 11-years solar cycle with variations lower than 15\%, which is a good model to describe the general behavior of the solar flux in a long-term period. On the other hand, for the geomagnetic indices prediction, it is recommended to use the cubic spline approach \citep{vallado_2005}. Machine Learning could potentially process this type of problem to describe the daily variations in the solar activity or in planetary amplitude. In this context, predicting the future behavior of the weather data with reasonable confidence is of particular interest. This challenging task has been already addressed by several authors, for instance, \citet{lean_2009} used a linear Autoregressive algorithm with lags based on the autocorrelation function. The highest correlations of each day is used to forecast the next one, which is similar to the simple naively forecasting method that we will use later in this work. \citet{henney_2012} used the global solar magnetic field to forecast the solar 10.7 cm (2.8 GHz) radio flux. A simple forecasting model is applied in \citet{warren_2017}, using a linear combination of the previous 81 observations to forecast the solar flux from 1 to 45 days. In this work, we apply Deep Learning methods for Time-Series forecasting, using historical data of solar activity (Since 1/10/1957 to 1/11/2021), available in the Earth Orientation Parameter (EOP) and Space Weather Data\footnote{\href{https://celestrak.com/SpaceData/}{https://celestrak.com/SpaceData/}, accessed on November 2021.}, to predict the behavior of the weather data and to calculate the atmospheric density.
   
\section{Methodology and Results}

\begin{figure}[!ht]
     \includegraphics[width=0.49\linewidth]{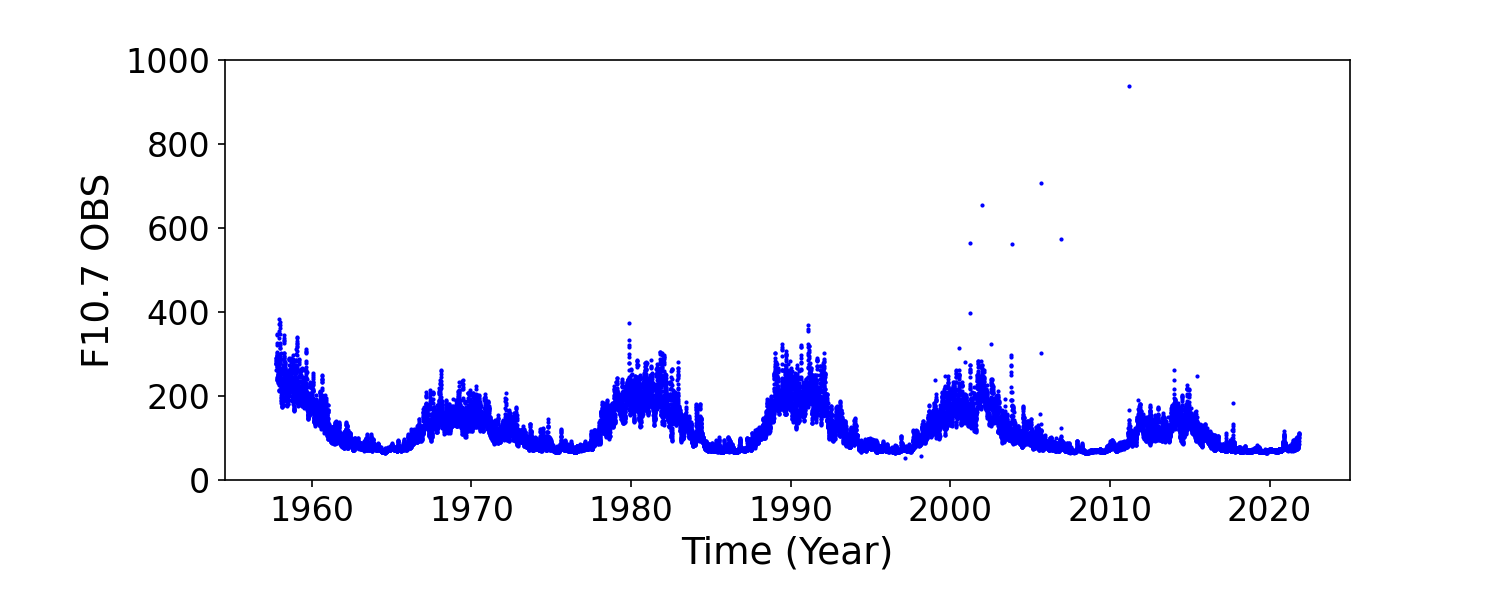}
     \includegraphics[width=0.49\linewidth]{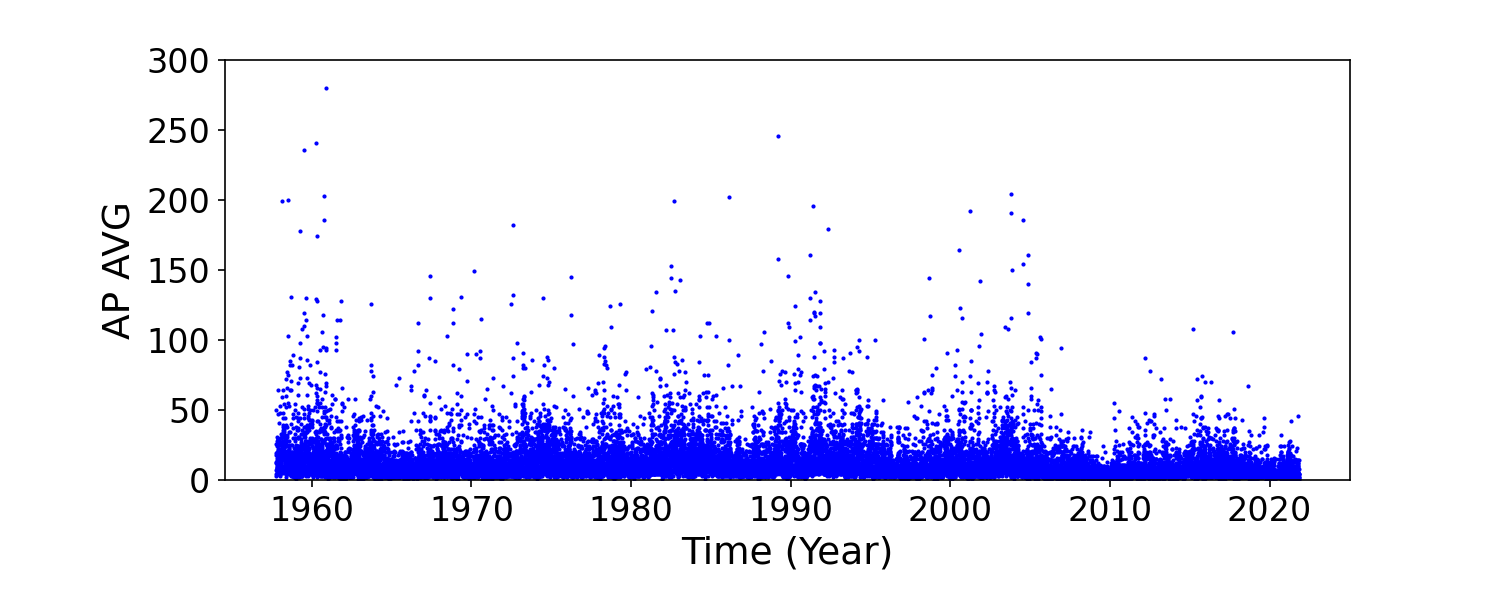}
    
      \caption{Space weather data vs. timestep from the EOP and Space Weather Data} \label{fig01_f10.7}
  \end{figure}

  Daily data, since 1/10/1957, for the Solar Radio Flux ({\bf F10.7 OBS}) is available in the Earth Orientation Parameter (EOP) and Space Weather Data. It is a univariate series, presented in the left panel of Fig. \ref{fig01_f10.7}. Observing this plot, we can notice that seasonality trends probably exists. Daily planetary amplitude ({\bf AP AVG}) is also available as an average of the 8 geomagnetic planetary amplitude, shown in the right panel of Fig. \ref{fig01_f10.7}. This parameter has integer values from 0 to 280, with only 102 observations with a value grater that 100.\\
  
\begin{figure}[!ht]
      \includegraphics[width=0.48\linewidth]{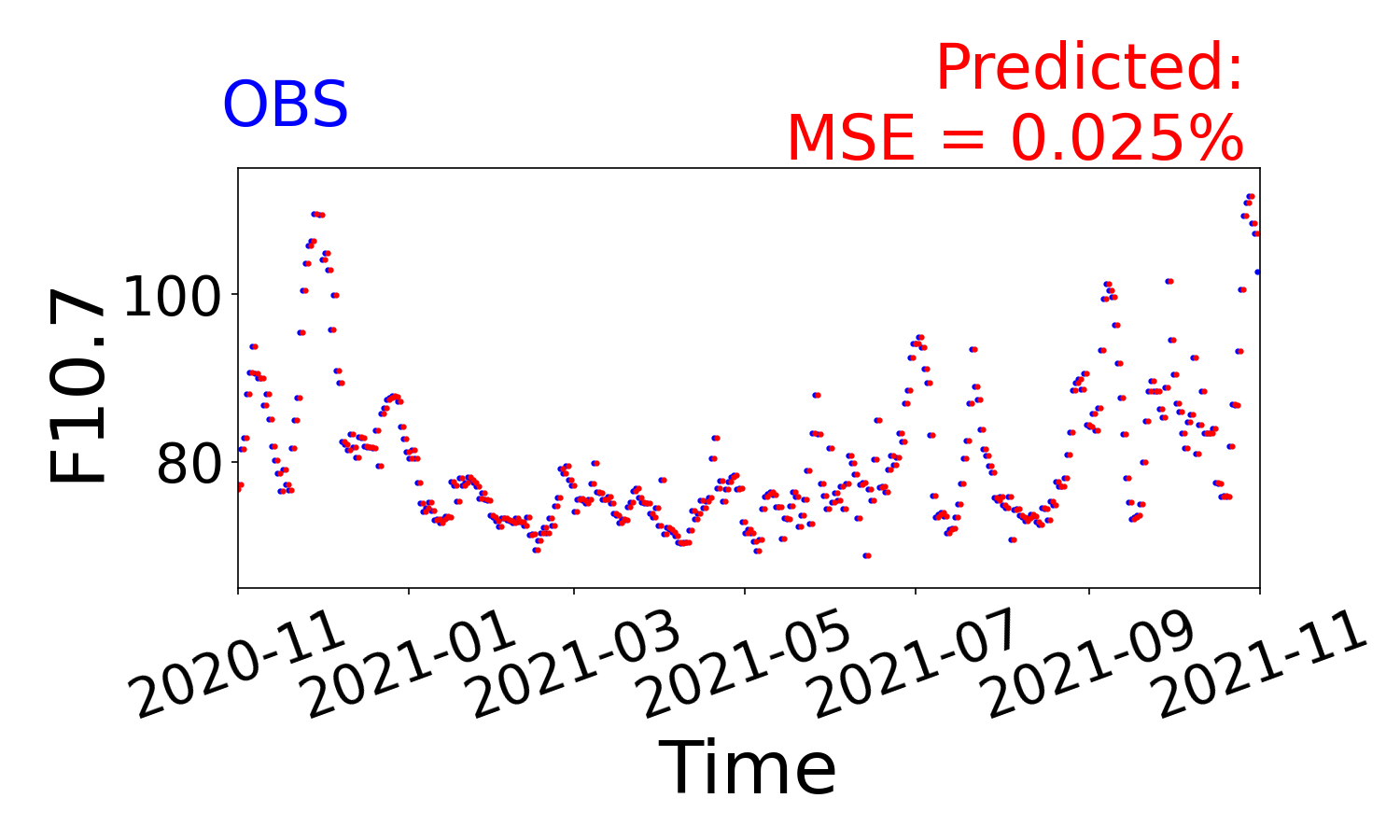}
      \includegraphics[width=0.48\linewidth]{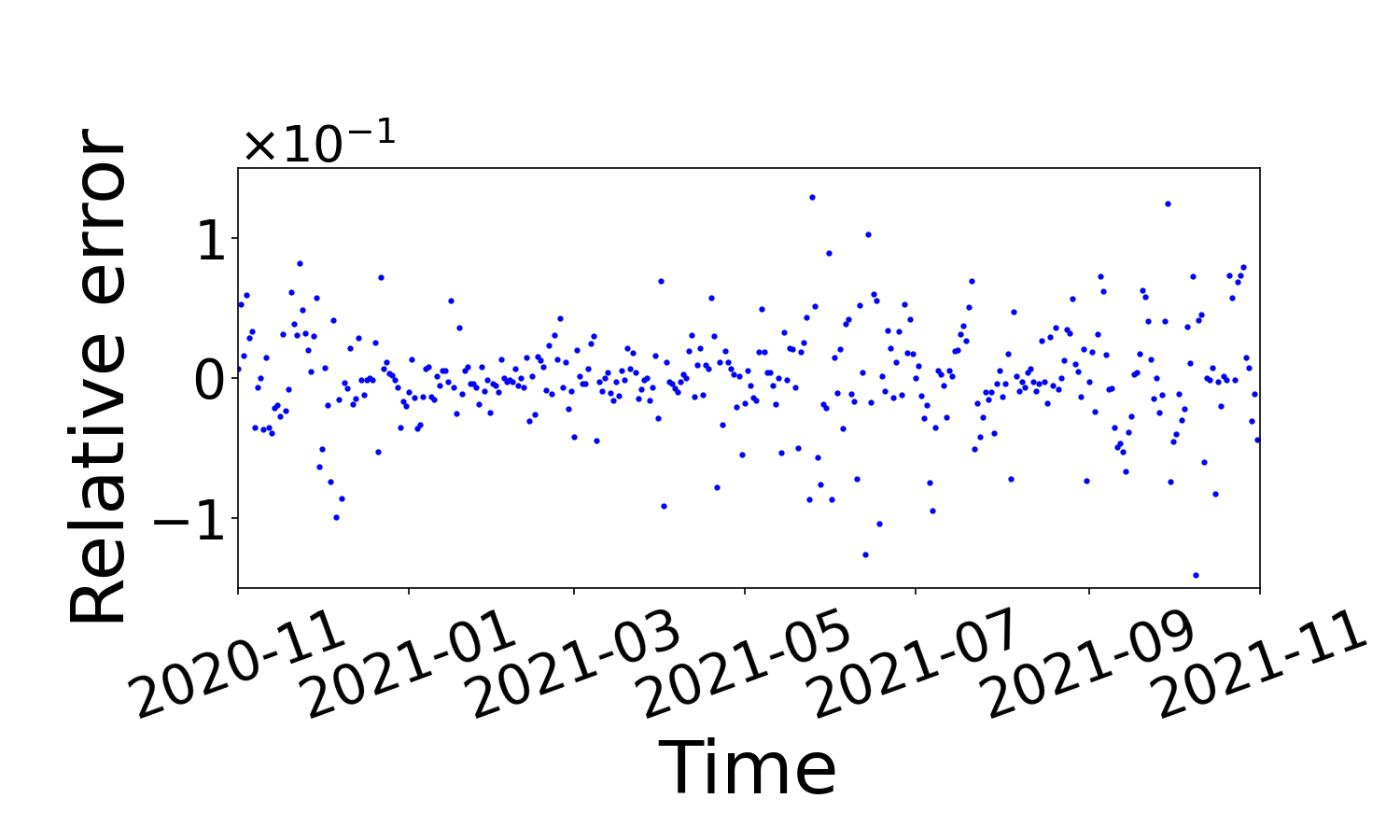}\\
      \includegraphics[width=0.48\linewidth]{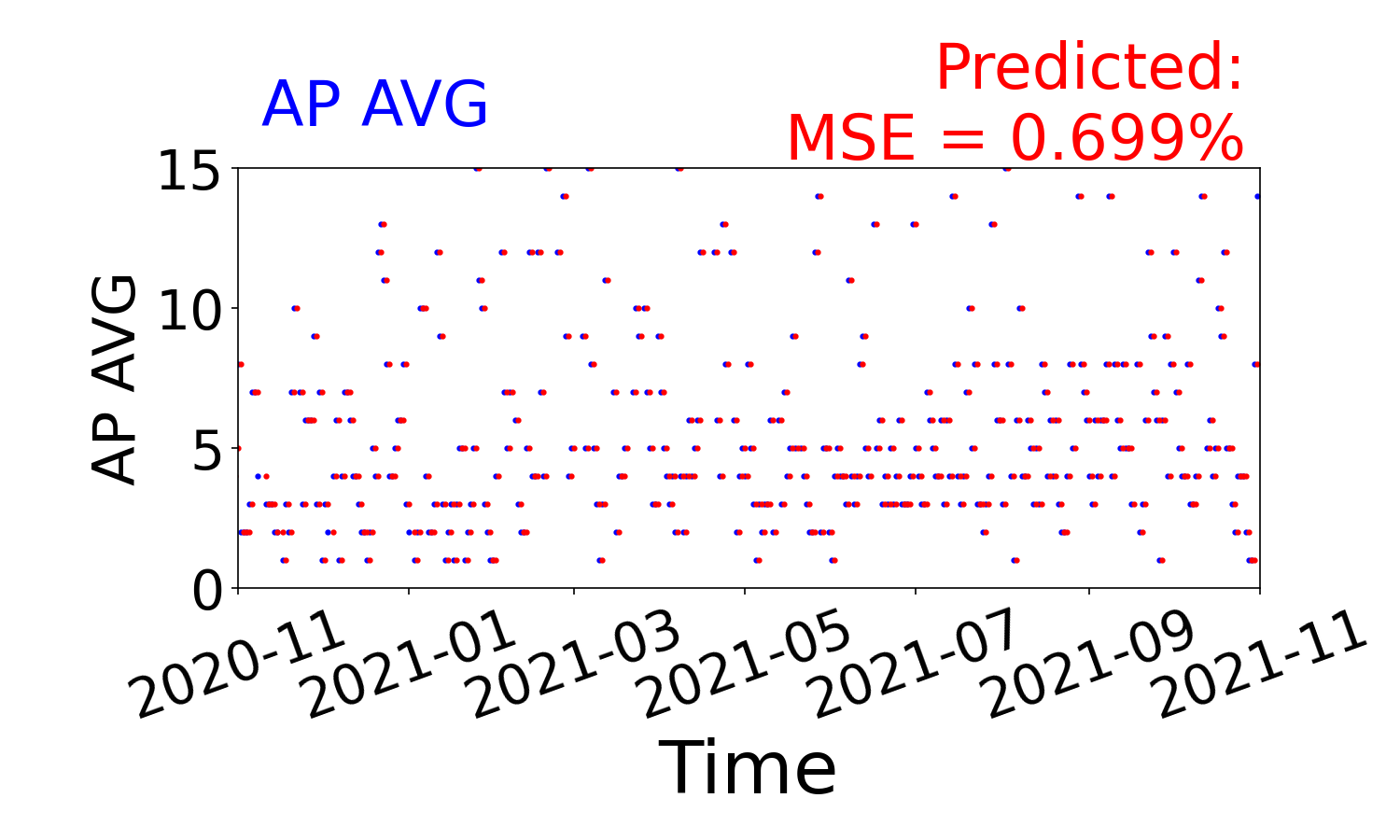}
      \includegraphics[width=0.48\linewidth]{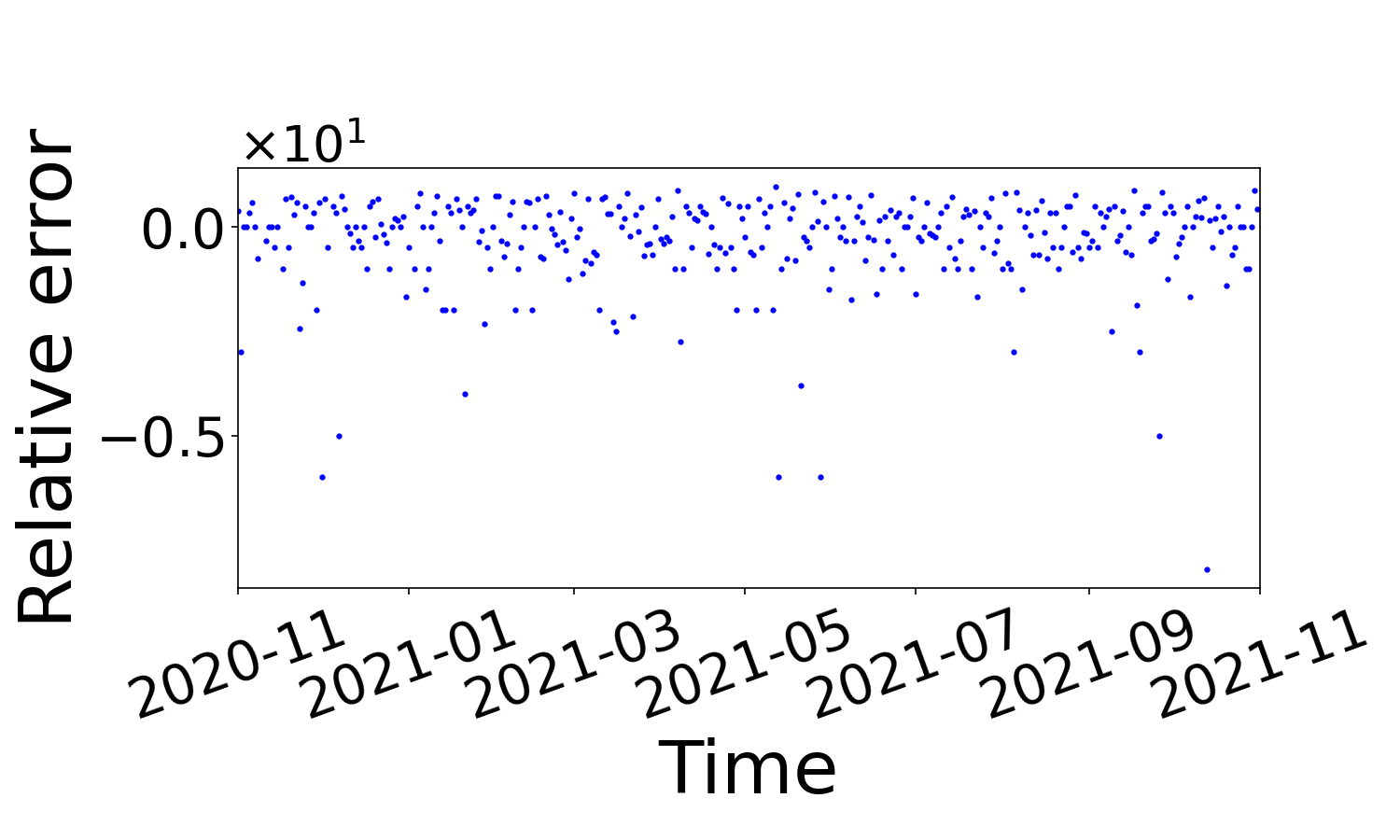}\\
      \caption{Naive one-step time series forecasting} \label{fig02_one_step_forecast}
  \end{figure}

    As a first step for a more complete study, we used Naive one-step time-series forecasting to predict the value of the parameters already mentioned, {\bf F10.7 OBS} and {\bf AP AVG}. The last 365 time steps (one year) are used as a test set to evaluate a very simple naively forecasting method, fitted on all the remaining observations. This strategy of forecasting simply teaks the previous period and applies it to the actual one. The walk-forward validation method is used to measure the performance of the model, where we applied one separate one-step forecast to each of the test observations. The true data was then added to the training set for the next forecast. We used the Mean absolute percentage error regression loss (MAPE) to compare our results with the real test set. This metric is defined in {\if \bf scikit-learn} documentation\footnote{\href{https://scikit-learn.org/stable/modules/model_evaluation.html\#mean-absolute-percentage-error}{https://scikit-learn.org/stable/modules/model\_evaluation.html\#mean-absolute-percentage-error}} as:
    \begin{eqnarray}\label{motion1}
       \text{MAPE} = \frac{1}{n}\sum_{i=0}^{n-1}\frac{|y_{i}-\hat{y}_{i}|}{max(\epsilon, |y_{i}|)}
    \end{eqnarray}
    \noindent where $y$ and $\hat{y}$ are the real and predicted value, respectively. $n$ is the number of the samples, and $\epsilon$ is an arbitrary small positive number to avoid undefined results when y is zero. Our results are presented in Fig. \ref{fig02_one_step_forecast}, where we can notice a good performance for {\bf F10.7 OBS} with an MAPE of 0.025\%,  and a relative error less than 0.15. A week performance is identified for the {\bf AP AVG} with an MAPE of 0.7\%, which is expected for the chaotic behavior of this parameter as shown in the right panel of Fig. \ref{fig01_f10.7}.\\
  
\begin{figure}[!ht]
      \includegraphics[width=0.48\linewidth]{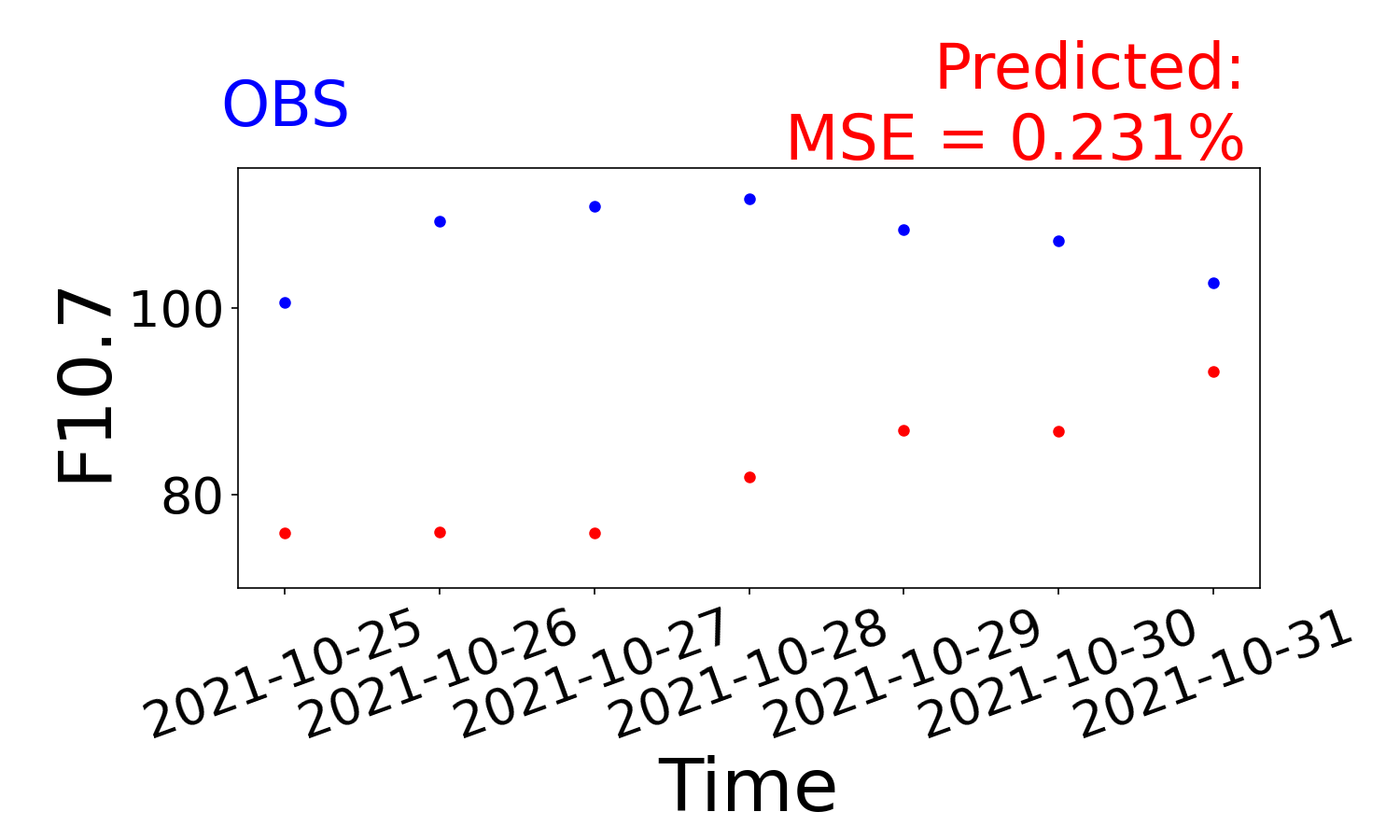}
      \includegraphics[width=0.48\linewidth]{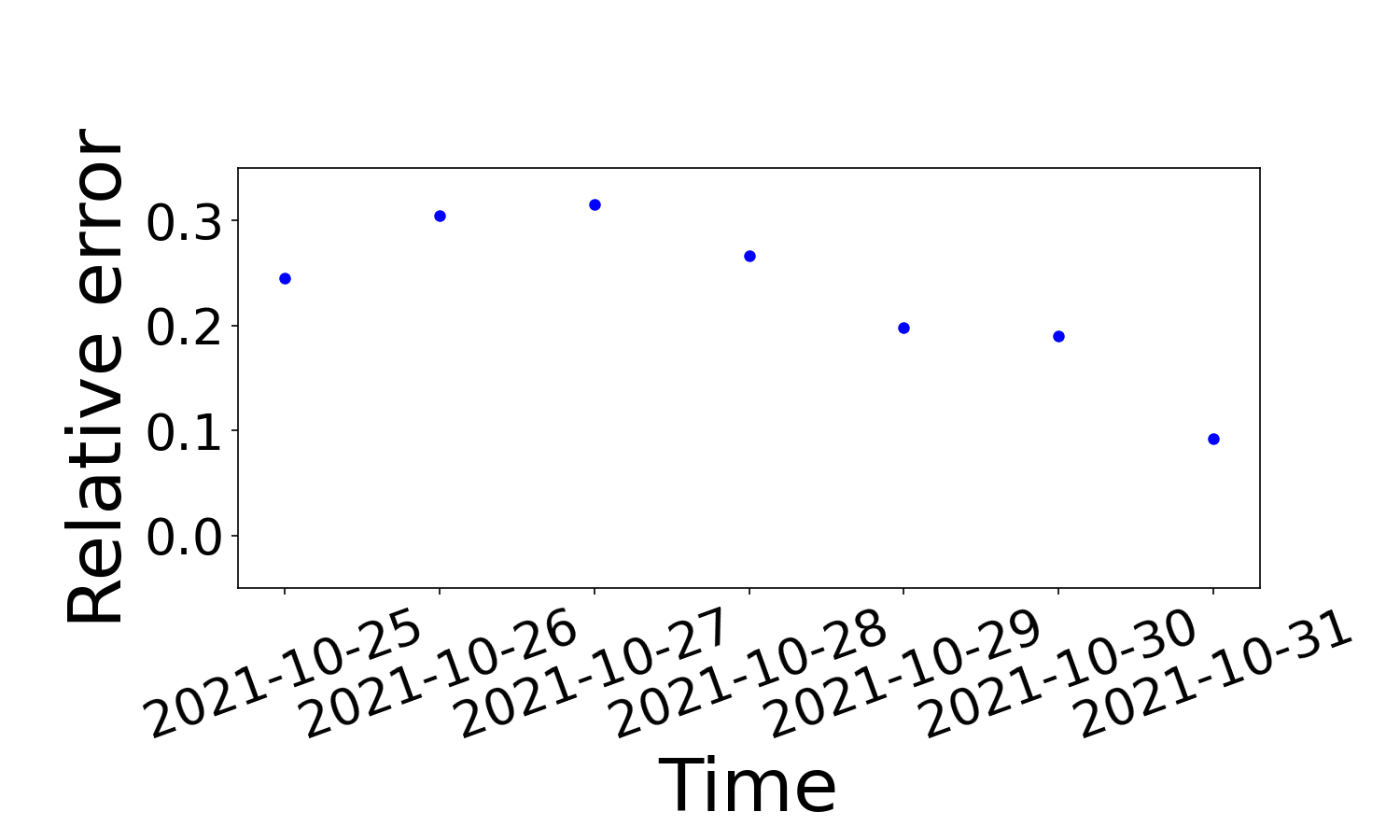}\\
      \includegraphics[width=0.48\linewidth]{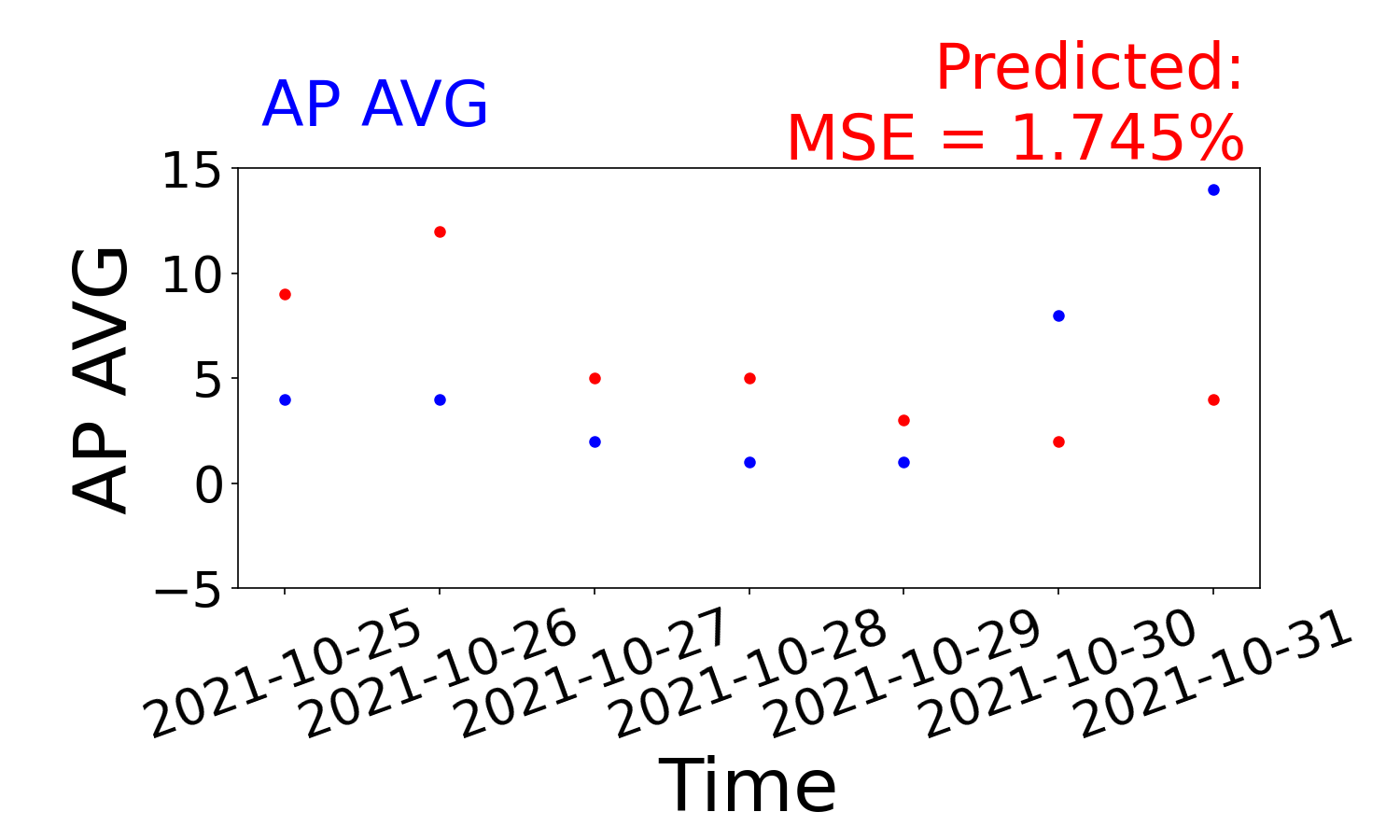}
      \includegraphics[width=0.48\linewidth]{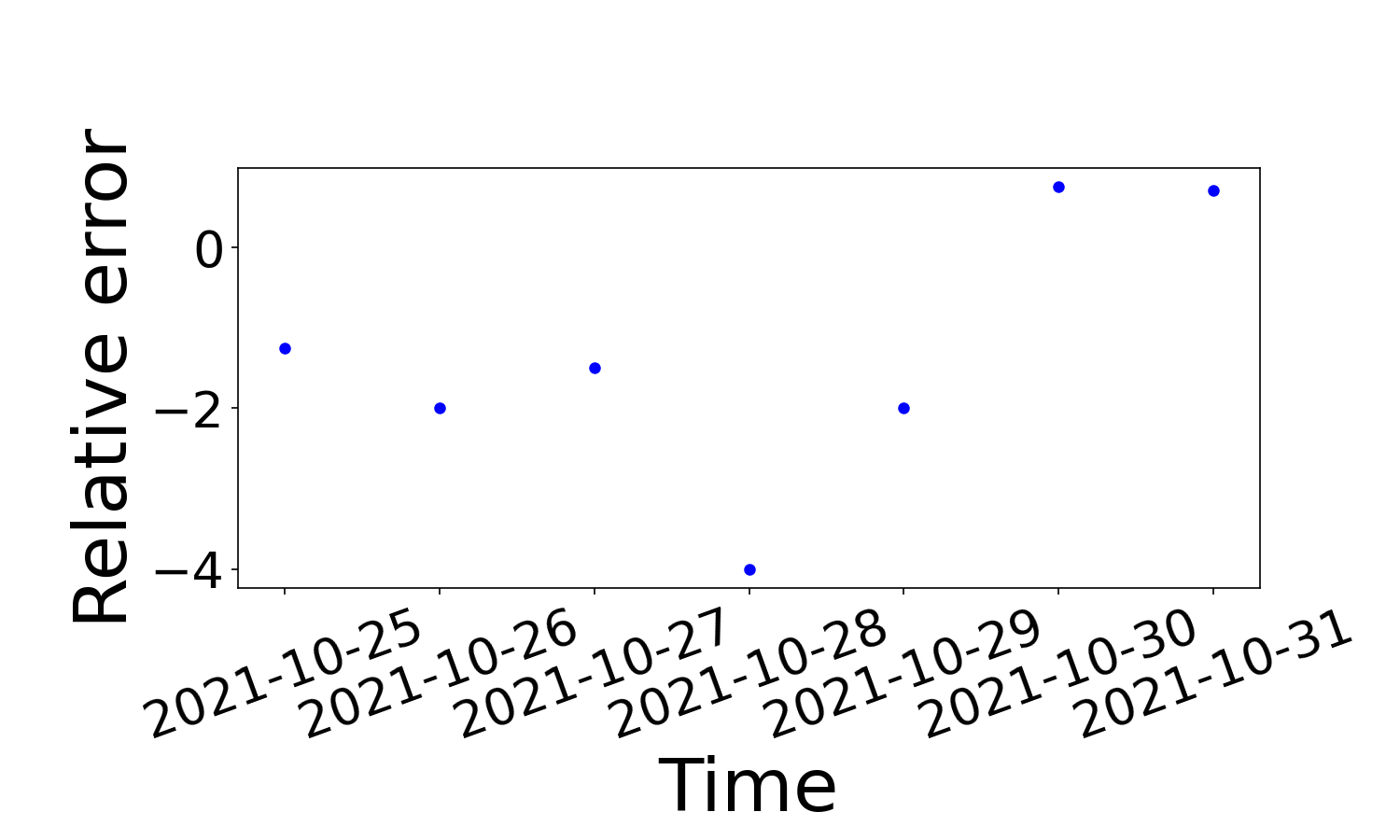}\\
      \caption{Naive One-week weather data Forecasting} \label{fig03_one_week_forecast_simple_model}
  \end{figure}

    Next, we tried to develop a deep learning models to make one-week forecasts. For this purpose, we first attempt to use a simple naive method, and then applied a Convolutional Neural Network (CNN). In the whole weather data, we have 23406 days, giving 3343 full weeks. We split the data into 3008 weeks as a train set and 335 as a test set. Here, we also used a walk-forward validation method to evaluate the models, where the model is used to predict one week, then the real data of this week is added to the training set. The process is repeated for all the weeks in the training set. Our results are presented in Fig. \ref{fig03_one_week_forecast_simple_model}, where we can notice a week performance with a MAPE of 0.231\%, and 1.745\%, and a relative error less than 0.3 and 6 for the selected parameters, {\bf F10.7 OBS} and {\bf AP AVG}, respectively.\\

displays the ACF of a regular particle in dynamical map for the Veritas family region. The right panel shows the ACF of a rather chaotic orbit
    
\begin{figure}[!ht]
      \includegraphics[width=0.48\linewidth]{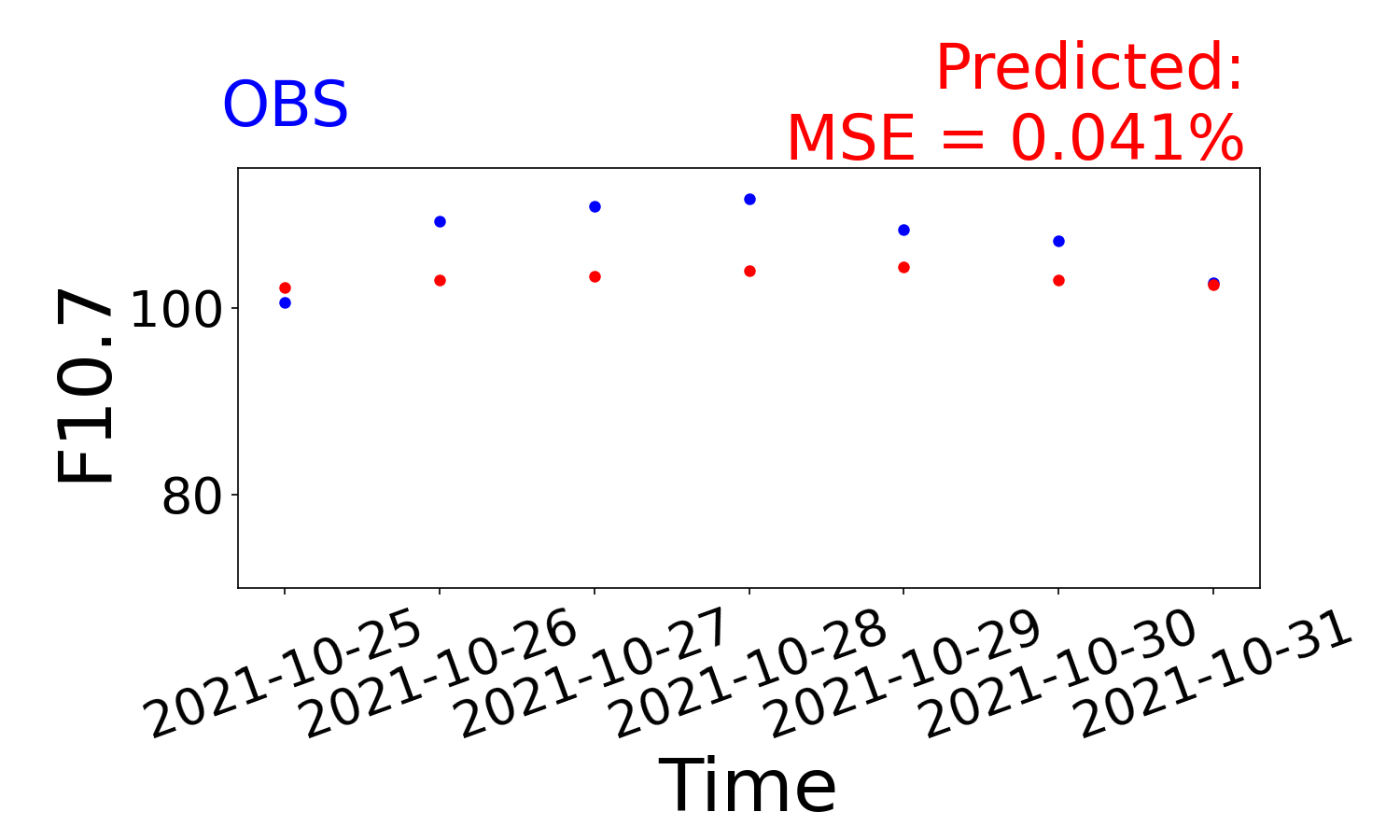}
      \includegraphics[width=0.48\linewidth]{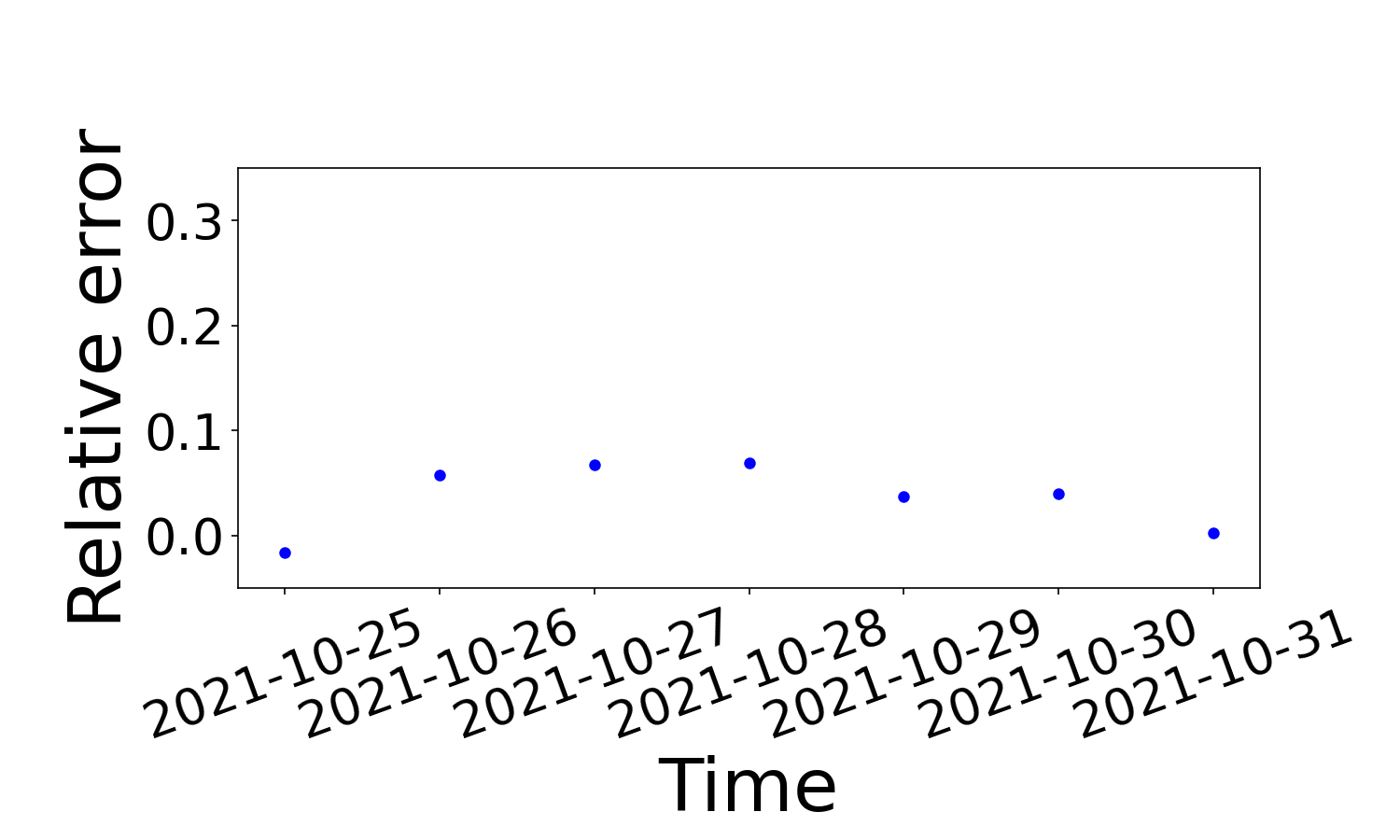}\\
      \includegraphics[width=0.48\linewidth]{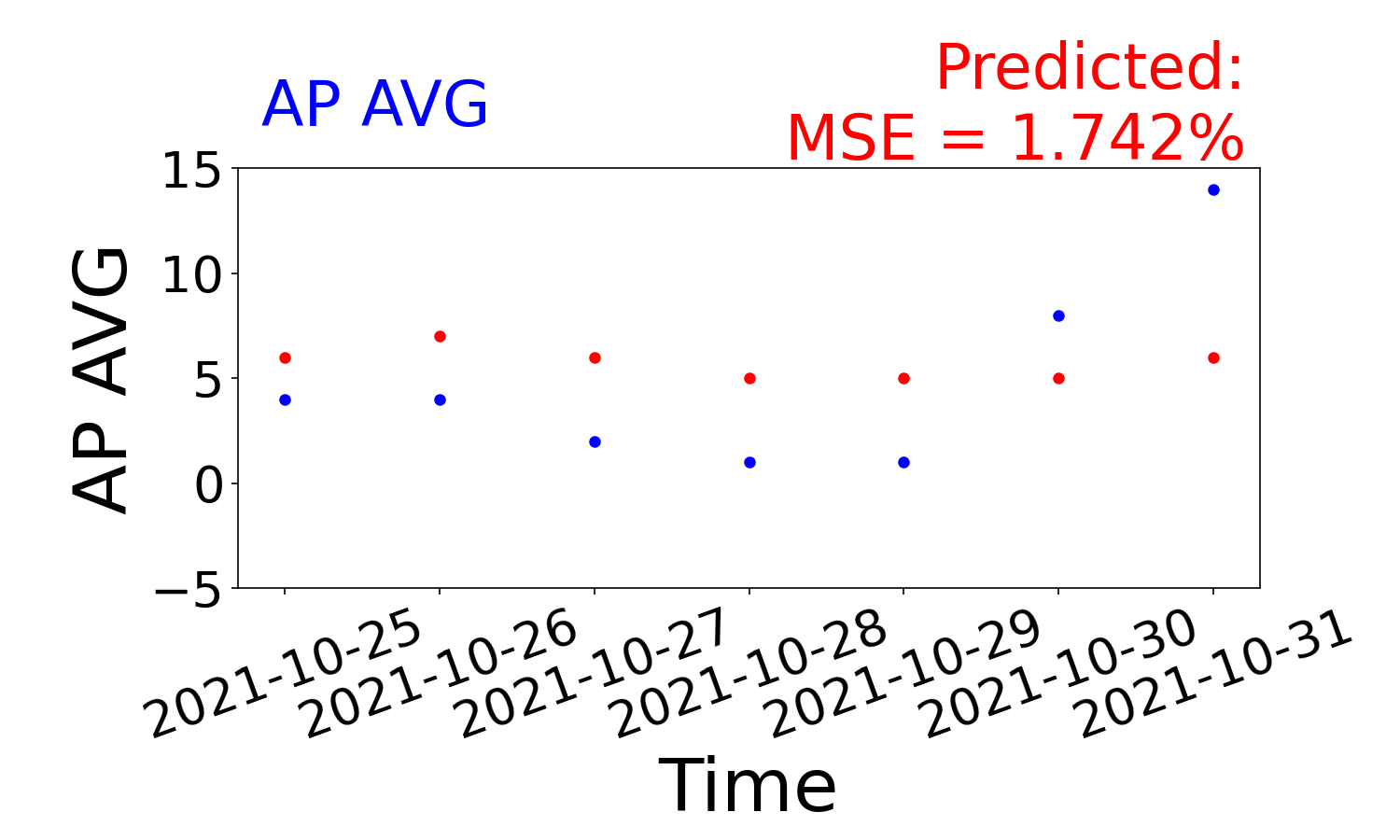}
      \includegraphics[width=0.48\linewidth]{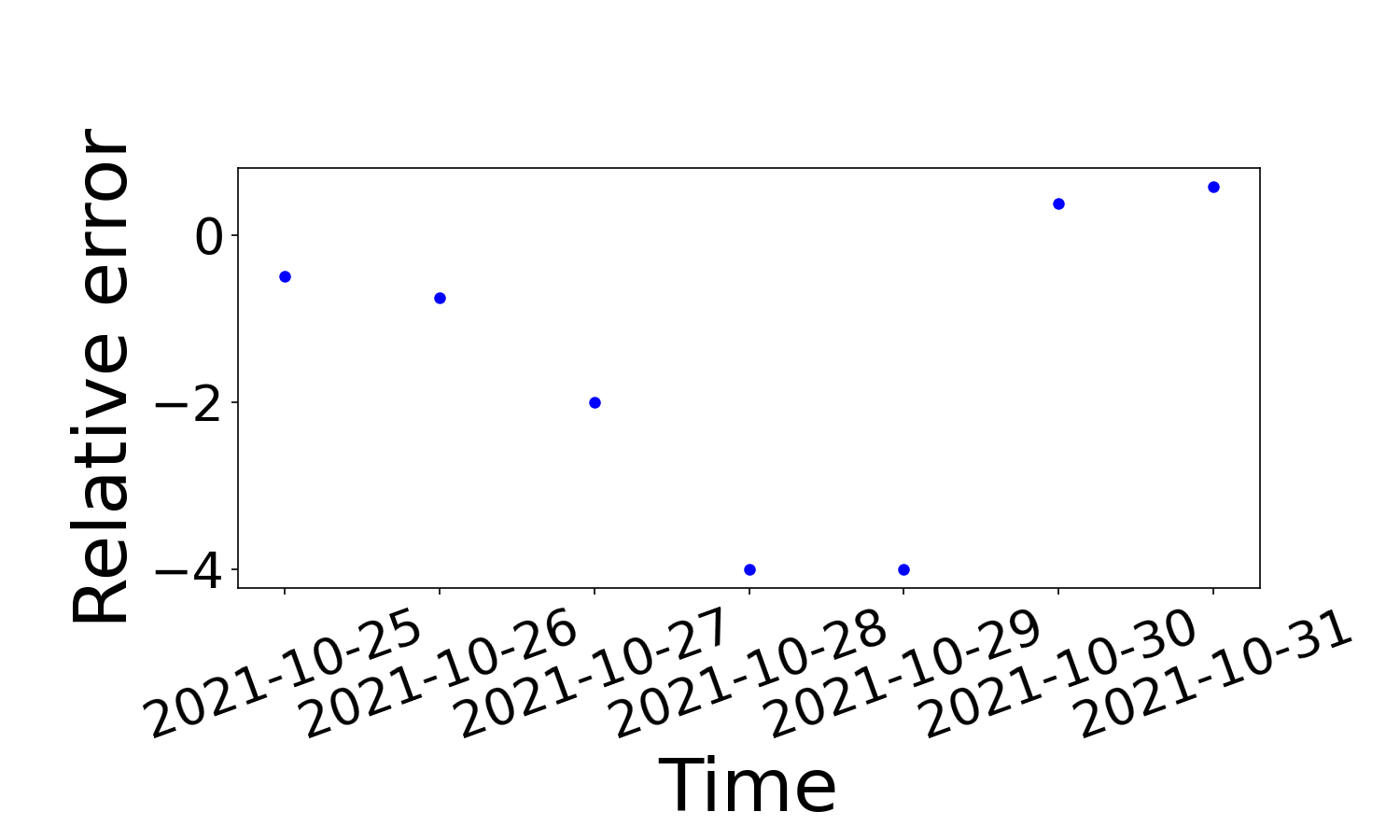}\\
      \caption{CNNs One-week weather data Forecasting} \label{fig04_one_week_forecast_cnn_model}
  \end{figure}

    To improve our results, we used a more sophisticated CNN method. CNN was originally created for image data \citep{pugliatti_2020, cassinis_2019, jianing_2022}, however, several CNN models can be adapted to time-series prediction tasks. In this work, we used a CNN model with two convolutional layers with the Rectified Linear Unit activation function (ReLU) defined to returm the positive part of the input. In each layer, a convolutional operation will read the intersted week 3 times (kernel size of 3) and the process will be performed 32 times (32 filters), then a pooling layer is added to select the maximum value over a window of size 2. The connected layers that interprets the features is then increased to 300 nodes. We fitted the model exposeing the model 100 times to the whole training dataset (100 epochs). The weights of the model are updated each epoch for each 12 samples (batch size of 12). For more details abour our process, we refer the reader to \citet{brownlee_2020}. Our results are presented in Fig. \ref{fig04_one_week_forecast_cnn_model}. We notice that the CNN can considerably reduce the MAPE for the solar flux ({\bf F10.7 OBS}) to 0.041\% and a relative error less than 0.1. However, the simple naive model provides very similar performance to the CNN model concerning the planetary amplitude ({\bf AP AVG}).\\
    
\section{Conclusion}
   In this work, we introduced a Machine Learning approach using a time series forecasting model to predict the daily variations in the solar activity and in planetary amplitude. This could be the first step for a more complete dynamical study to reduce the uncertainly of the Drag due to the estimation of the atmospheric density. We used the historical data of solar activity from 1957 to the present day from the EOP and Space Weather Data. A good agreement was found between the predicted and real data. We applied a simple naive method, and a Convolutional Neural Network (CNN). The walk-forward validation method is used in order to make the best possible forecast at each time step. In this method, we train the model as new data becomes available. A relative error less than 0.15 with respect to the analytical model of the solar flux is reached making a single one-step prediction, while the same model provides a relative error less than 0.3 making a single one-week prediction. However, the CNN model reduced this error to less than 0.1. A more complete study is necessary to reduce the error in the prediction of the geomagnetic indices.
\section*{Acknowledgments}

   The authors would like to thank the Institutional Training program (PCI/INPE), which supported this work via the grant (444327/2018-5), São Paulo Research Foundation (Fapesp) project number 2016/24561-0, and Coordination for the Improvement of Higher Education Personnel (CAPES) project PrInt CAPES-INPE. 
   

%

%
%

\end{document}